%% file: main.tex
\documentclass[10pt,conference]{IEEEtran}
\IEEEoverridecommandlockouts
\usepackage{cite}
\usepackage{amsmath,amssymb,amsfonts, mathrsfs}
\usepackage{algorithmic}
\usepackage{graphicx}
\usepackage{textcomp}
\usepackage{xcolor}

\usepackage{amsmath,amsfonts,amssymb, mathrsfs}
\usepackage{bigints}
\usepackage{subfig}
\usepackage{graphicx}
\usepackage[colorlinks=true, allcolors=blue]{hyperref}
\usepackage{stmaryrd}

\usepackage[braket,qm]{qcircuit}
\usepackage{braket}
\usepackage{physics}

\usepackage{multicol}

\def\BibTeX{{\rm B\kern-.05em{\sc i\kern-.025em b}\kern-.08em
    T\kern-.1667em\lower.7ex\hbox{E}\kern-.125emX}}

\begin{document}

\title{Comparison of Amplitude Estimation Algorithms by Implementation}

\author{\IEEEauthorblockN{Kwangmin Yu}
\IEEEauthorblockA{\textit{Computational Science Initiative} \\
\textit{Brookhaven National Laboratory}\\
Upton, New York 11973, USA \\
kyu@bnl.gov}
\and
\IEEEauthorblockN{Hyunkyung Lim}
\IEEEauthorblockA{\textit{Department of Applied Math. and Stat.} \\
\textit{Stony Brook University}\\
Stony Brook, New York 11794, USA \\
hyun-kyung.lim@stonybrook.edu}
\and
\IEEEauthorblockN{Pooja Rao}
\IEEEauthorblockA{\textit{Department of Mathematics} \\
\textit{Stony Brook University}\\
Stony Brook, New York 11794, USA \\
pooja.rao@stonybrook.edu}
\and
\IEEEauthorblockN{Dasol Jin}
\IEEEauthorblockA{\textit{Department of Applied Math. and Stat.} \\
\textit{Stony Brook University}\\
Stony Brook, New York 11794, USA \\
dasol.jin@stonybrook.edu}

}

\maketitle

\begin{abstract}

Since the quantum amplitude estimation (QAE) was invented by Brassard et al., 2002, several advanced algorithms have recently been published (Grinko et al., 2019, Aaronson et al, and Suzuki et al., 2020). The main difference between the variants and the original algorithm is that the variants do not need quantum phase estimation (QPE), a key component of the canonical QAE (Brassard et al., 2002), that is composed of many expensive operations on NISQ devices. In this paper, we compare and analyze two of these new QAE approaches (Grinko et al., 2019, and Suzuki et al., 2020) by implementation using the Qiskit package. The comparisons are drawn based on number of oracle queries, quantum circuit depth, and other complexities of implementation for a fixed accuracy. We discuss the strengths and limitations of each algorithm from a computational perspective.

\end{abstract}


\input{introduction}

\input{quantum_couting_algs}

\input{implementation}

\input{results}

\input{conclusion}

\bibliography{report}
\bibliographystyle{IEEEtran}

\input{appendix}


\end{document}

%% file: introduction.tex
\section{Introduction}
\label{sec:introduction}

Unprecedented development has taken place in recent years in a range of quantum computing algorithms.
In particular, significant progress has been made in building quantum computers by companies, such as IBM, Google, and Rigetti. 
Recently, Google, IBM, and Intel have announced 72 qubits \cite{Kelly2018Google}, 50 qubits \cite{Knight2017IBM} (53 qubits \cite{Shankland2019IBM}), and 49 qubits \cite{Hsu2019Intel} quantum devices, respectively. 
More notably, IBM has made available its cloud enabled quantum computing platform to the public.  
These currently available Noisy Intermediate-Scale Quantum (NISQ) devices provide a tangible quantum programming environment, and are serving as a stepping stone for the large-scale universal quantum computers of the future \cite{Preskill2018quantumcomputingin}. 

The quantum amplitude estimation (QAE) by Brassard et al. \cite{brassard2002quantum} is one of the most notable quantum algorithms that has advanced not just from the viewpoint of implementation on NISQ devices, but also in terms of better efficiency (required number of qubits, quantum circuit depth)~\cite{grinko2019iterative, suzuki2020amplitude, aaronson2020quantum, wie2019simpler, nakaji2020faster}.
The canonical QAE (Brassard et al.) is a monumental quantum algorithm and a brilliant extension of Grover's algorithm~\cite{grover1996fast} in conjunction with Quantum Amplitude Amplification (QAA).
However, it is extremely challenging and in some cases infeasible to implement on NISQ devices, because of the required number of controlled operations of Quantum Phase Estimation (QPE) \cite{Yu2020Practical}.
The common and most crucial characteristic of the recent QAE algorithms is their lack of use of QPE, which makes them better suited for the near term quantum devices.


In this study, we implement and compare two recent QAEs concomitant with QAA on $10$ and $14$-qubit search domains using Qiskit \cite{qiskit}.
The first by Grinko et al. \cite{grinko2019iterative} (IQAE) uses iterative optimization of QAE and the second by Suzuki et al.~\cite{suzuki2020amplitude} (MLQAE) uses the maximum likelihood estimation after making measurements on several circuits.
We compare these two QAEs from the perspective of accuracy and efficiency for quantum Monte Carlo integration.
Even though the QAE by Aaronson et al. \cite{aaronson2020quantum} has the same asymptotic complexity with the others, we did not include it in our analysis because of the large constant factor associated with the QAE (Fig. 3 in~\cite{grinko2019iterative}), rendering it impractical on NISQ devices for our domain size ($2^{10}$ and $2^{14}$).

\section{Notations}

In the following sections, Dirac's bra-ket notation will be used for qubit representation and arithmetic, such as the tensor product.
For a multiple qubit system, consecutive binary number strings have the most significant qubit located on the left and the least significant qubit on the right.
For example, we have $\ket{0} \otimes \ket{1} \otimes \ket{1} = \ket{0} \ket{1} \ket{1} = \ket{011}$ in the binary representation, and $\ket{011} = \ket{3}$ and $\ket{110} = \ket{6}$ in the decimal representation of the computational basis.
In the decimal representation, the number of qubits $n$ is denoted by a subscript as in $\ket{0}_n$.
The dimension of a square matrix is also denoted by a subscript.
For example, $\mathbb{I}_{n}$ denotes the $n\times n$ identity matrix.

In quantum circuit diagrams, the top and the bottom qubits represent the least and the most significant qubits, respectively. 
The Hadamard matrix, H = $\frac{1}{\sqrt{2}} \begin{pmatrix} 1 & 1\\ 1 & -1\end{pmatrix}$, is a unitary matrix that is also Hermitian, so it is its own inverse.
The three Pauli matrices (Pauli gates when they are used in quantum circuits) $X$, $Y$ and $Z$ are as follows:

\begin{equation*}
\label{eq:pauli}
    X = \begin{pmatrix} 0 & 1\\ 1 & 0\end{pmatrix}, ~~
    Y = \begin{pmatrix} 0 & -i\\ i & 0\end{pmatrix}, ~~
    Z = \begin{pmatrix} 1 & 0\\ 0 & -1\end{pmatrix}.
\end{equation*}



%% file: quantum_couting_algs.tex
\section{Quantum Amplitude Estimation}
\label{sec:qae}


In this section, we briefly review quantum amplitude amplification (QAA) and quantum amplitude estimation (QAE) algorithms  \cite{brassard2002quantum}. The quantum amplitude amplification is a generalization of Grover's search algorithm \cite{grover1996fast}, without the loss of the quadratic quantum speedup over its classical counterpart.
While the original Grover's algorithm searches one solution in the given domain, QAA searches multiple solutions in the given domain.
When the number of solutions is of more interest than an individual solution, QAE is applied to estimate the number of solutions.
QAA and QAE are fundamental building blocks for quantum implementation of hit or miss Monte Carlo integration \cite{Yu2020Practical}.

Suppose we have a boolean function $f$ and a unitary operator $\mathcal{A}$, where $f$ maps the good states to $\ket{1}$ and the bad states to $\ket{0}$ on domain $\mathscr{D}$ such that $| \mathscr{D} | = N = 2^n$, and $\mathcal{A}$ acts on $n+1$ qubits such that
\begin{equation}
\label{eq:qae_a}
    \ket{\Psi} = \mathcal{A} \ket{0}_n \ket{0} = \sqrt{1-a} \ket{\psi_0}_n \ket{0} + \sqrt{a} \ket{\psi_1}_n \ket{1},
\end{equation}
\noindent where the good state is $\ket{\psi_1}_n$ with $|\ket{\psi_1}_n| = k$, the bad state is $\ket{\psi_0}_n$, and we have unknown $a = k / N \in [0,1]$.
The job of QAE is to find $a$ approximately.
Since $\mathcal{A}$ knows the solutions and classifies them into $\ket{1}$ (good states) and $\ket{0}$ (bad states) in the last qubit in Eq.~(\ref{eq:qae_a}), $\mathcal{A}$ is a quantum oracle.
The algorithm complexity is measured by the number of quantum queries to the operator $\mathcal{A}$.

To achieve the quantum speedup, instead of measuring the last qubit of $\ket{\Psi} = \mathcal{A} \ket{0}_n \ket{0}$ directly, $\ket{\Psi}$ is first amplified by the following unitary operator $\textbf{Q}$:

\begin{equation}
\label{eq:qae_def_q}
    \textbf{Q} = \mathcal{A} \textbf{S}_0 \mathcal{A}^{-1} \textbf{S}_{\chi},
\end{equation}

\noindent where $\textbf{S}_0 = \mathbb{I}_{n+1} - 2 \ket{0}_{n+1} \bra{0}_{n+1}$ and
$\textbf{S}_{\chi} = ( \bigotimes\limits^{n} \mathbb{I}_2 ) \otimes Z$.
$\textbf{S}_{\chi}$ puts a negative sign to the good state $\ket{\psi_1}_n \ket{1}$ and does nothing to the bad state $\ket{\psi_0}_n \ket{0}$.
Let us define a parameter $\theta \in [0, \pi/2]$ so that $\sin^2 \theta = a$. With this, we can rewrite Eq.~(\ref{eq:qae_a}) as:
\begin{equation}
\label{eq:qae_theta}
    \ket{\Psi} = \mathcal{A} \ket{0}_n \ket{0} = \cos \theta \ket{\psi_0}_n \ket{0} + \sin \theta \ket{\psi_1}_n \ket{1}.
\end{equation}

\noindent By applying $\textbf{Q}$ (amplitude amplification operator) repeatedly $m$ times on $\ket{\Psi}$, we get 

\begin{equation}
\label{eq:qae_qm}
    \textbf{Q}^m \ket{\Psi} = \cos ((2m+1) \theta) \ket{\psi_0} \ket{0} + \sin ((2m+1) \theta) \ket{\psi_1} \ket{1}.
\end{equation}

\noindent From Eqs.~(\ref{eq:qae_a}) and (\ref{eq:qae_theta}), it can be observed that the measurement after applying $\textbf{Q}^m$ on $\mathcal{A} \ket{0}_n \ket{0}$ shows a quadratically larger probability of obtaining the good state (provided $\theta$ is sufficiently small so that $ (2m+1) \theta < \frac{\pi}{2}$) than measuring $\mathcal{A} \ket{0}_n \ket{0}$ directly \cite{brassard2002quantum}.


The canonical QAE~\cite{brassard2002quantum} estimates $\theta$ in Eq.~(\ref{eq:qae_theta}) by QPE which includes the inverse QFT. QPE is implemented by the controlled operation on $\textbf{Q}$ operator. 
Hence, it needs a number of multi-controlled operations, which are further decomposed into many basis gates, increasing the circuit depth.
When this algorithm is implemented on NISQ devices, the accuracy is strongly limited by the lack of direct connectivity between qubits because all the ancilla registers need connectivity with the target register and a sufficiently large number of ancillae are needed to ensure desired accuracy of the estimation.   

On the other hand, MLQAE \cite{suzuki2020amplitude} is implemented by post-processing the result of the quantum computation (and measuring the process without QPE) using maximum likelihood estimation. Since different levels of amplification (depending on power of $\textbf{Q}$) can be executed independently, the algorithm is parallelizable. 
IQAE \cite{grinko2019iterative} also post-processes results from quantum circuit runs. In the post-processing, it estimates the optimal power $k$ of $\textbf{Q}$ and applies $\textbf{Q}^k$ to $\ket{\Psi}$ (in Eq. (\ref{eq:qae_a})) in the next iteration. The iteration continues until the specified error bound is met.
The main benefit of MLQAE and IQAE over the canonical QAE is that these variants do not need ancilla qubits to read out the amplitude $\sqrt{a}$ of Eq.~(\ref{eq:qae_a}).
The analysis of error bounds and the number of queries (application of $\mathcal{A}$ and $\mathcal{A}^{-1}$) of MLQAE and IQAE are discussed in \cite{suzuki2020amplitude, grinko2019iterative, aaronson2020quantum}.
Both MLQAE and IQAE need $O \bigl( \frac{1}{\epsilon} \bigl( \sqrt{\frac{K}{N}} \bigr)   \bigr)$ oracle queries to estimate $K$ good states in $N$ samples (domain) with error $\epsilon$ \cite{aaronson2020quantum}.
The implementation aspects and results obtained from executing MLQAE and IQAE on quantum simulator will be discussed in the following sections.



%% file: implementation.tex
\section{Implementation}
\label{sec:implementation}

\begin{figure}[t]
\centering
  \subfloat[$\mathcal{A}$]{%
    \includegraphics[width=.24\textwidth]{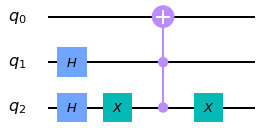}}
  \subfloat[$\mathcal{A}^{-1}$]{%
    \includegraphics[width=.24\textwidth]{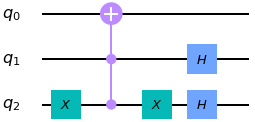}
  }\\
  \caption{Quantum circuit implementation of $\mathcal{A}$ from Eq.~(\ref{eq:qae_a}) in Qiskit.}\label{fig:A}
\end{figure}

\begin{figure*} [ht]
\begin{center}
\includegraphics[width=0.99\linewidth]{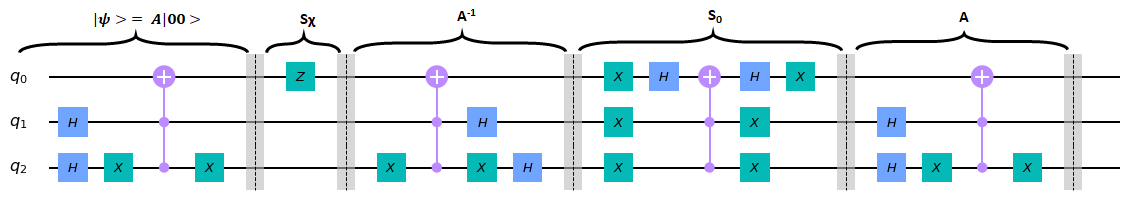}
\end{center}
\caption{Quantum circuit implementation of $\mathcal{A}$ and $\textbf{Q}$ of $2$-qubit domain from Eqs.~(\ref{eq:qae_a}) and~(\ref{eq:qae_def_q}), in Qiskit.}
\label{fig:Q} 
\end{figure*}

\begin{figure*} [ht]
\begin{center}
\includegraphics[width=0.99\linewidth]{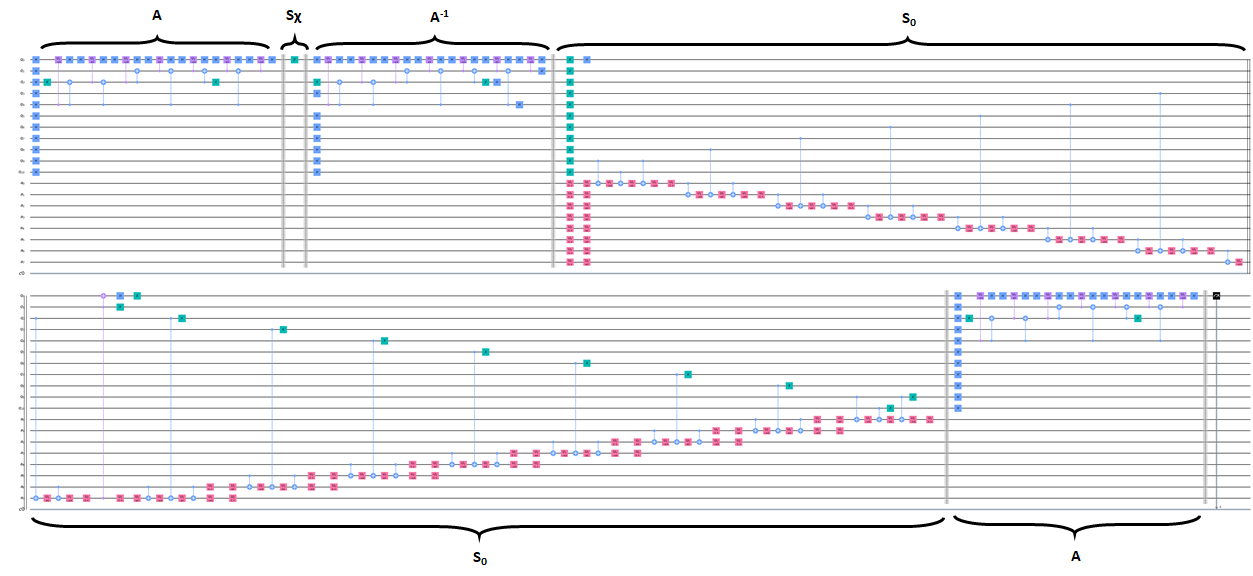}
\end{center}
\caption{Quantum circuit implementation of $\mathcal{A}$ and $\textbf{Q}$ of $10$-qubit domain from Eqs.~(\ref{eq:qae_a}) and~(\ref{eq:qae_def_q}), in Qiskit.}
\label{fig:Q_10q} 
\end{figure*}

Fig.~\ref{fig:A} shows an example of $\mathcal{A}$ and $\mathcal{A}^{-1}$ which have the solution $\ket{01}$ (good state) out of the four possible values, $\ket{00}$, $\ket{01}$, $\ket{10}$, and $\ket{11}$ in a $2$-qubit domain.
The operator $\textbf{Q}$ in Eq.~(\ref{eq:qae_def_q}) is applied on $\mathcal{A}$ (in Fig.~\ref{fig:A}) as depicted in Fig.~\ref{fig:Q}. 
Fig.~\ref{fig:Q_10q} shows $\textbf{Q} \mathcal{A}$ implementation on a $10$-qubit domain.
Both MLQAE and IQAE commonly run $\textbf{Q}^k \mathcal{A}$ circuits several times and conduct post-processing based on measurement results of the circuits.
The main difference lies in how each method chooses the power $k$ of $\textbf{Q}$ and what specific post-processing technique it employs.



\subsection{Maximum Likelihood Quantum Amplitude Estimation}

When only one circuit (only $\mathcal{A}$) is used, then there is no quantum speedup~\cite{suzuki2020amplitude}. More circuits increase the accuracy of the estimation of the amplitude. 
Suzuki et al.~\cite{suzuki2020amplitude} discuss two options for circuit sequencing in MLQAE, linearly incremental sequence (LIS) and exponential incremental sequence (EIS), and suggest EIS as the asymptotically optimal choice.
Thus, we adapt the EIS which has exponential power of $\textbf{Q}$ from the second circuit. For example, the $n$-th circuit has $\textbf{Q}^{2^{n-2}}$ operator applied to it following the application of $\mathcal{A}$ for $n > 1$.


The key idea of MLQAE is post-processing of measurements from each of the quantum circuits.
The likelihood functions and the resultant maximum likelihood function are defined as following in the domain $[0, \pi/2]$ for $\theta$:
\begin{equation}
\label{eq:lfunc}
    L_k (h_k ; \theta) = \lbrace \sin^2 ((2m_k+1) \theta) \rbrace^{h_k} \lbrace \cos^2 ((2m_k+1) \theta) \rbrace^{N - h_k},
\end{equation}

\begin{equation}
\label{eq:mlfunc}
    L (\vec{\textbf{h}} ; \theta) = \prod\limits_{k=0}^{m}  L_k (h_k ; \theta),
\end{equation}

\noindent where $m_k$, $h_k$, and $N$ are the power of $\textbf{Q}$, hit count of $1$, and number of shots for the $k$-th circuit, respectively, and $\vec{\textbf{h}} = (h_0, h_1, \cdots, h_m)$.
The maximum likelihood estimation estimates $\hat{\theta}$, which maximizes $L (\vec{\textbf{h}} ; \hat{\theta})$ in the domain.
But instead of $L (\vec{\textbf{h}} ; \theta)$, $\log L (\vec{\textbf{h}} ; \theta)$ is used to estimate $\hat{\theta}$ since the $\log$ function is monotonically increasing.

\subsection{Iterative Quantum Amplitude Estimation}

IQAE \cite{grinko2019iterative} also utilizes post-processing to estimate optimal power $k$ of $\textbf{Q}$ at each iteration.
It starts from $k=0$ ($\ket{\Psi} = \mathcal{A} \ket{0}_n \ket{0}$) with confidence interval $[\theta_l, \theta_u] \subseteq [0, \pi/2]$ where $\theta_u$ and $\theta_l$ are upper and lower bounds of $\theta$, respectively, in Eqs.~(\ref{eq:qae_theta}) and~(\ref{eq:qae_qm}).
The post-processing searches the maximum $k$ which puts $[(4k + 2) \theta_l, (4k + 2) \theta_u]$ in $[0, \pi]$ (upper half-plane) or $[\pi, 2\pi]$ (lower half-plane).
This method derives from the fact that the measurement probability of the good state in Eq.~(\ref{eq:qae_qm}) is $\sin^2 ((2k+1) \theta) = \{1 - \cos ((4k+2) \theta) \} / 2$ when the power of $\textbf{Q}$ is $k$ and that a cosine function is monotonic in the upper half-plane and the lower half-plane.

%% file: results.tex
\section{Results}
\label{sec:results}


\begin{figure*}[t]
\centering
  \subfloat[$m=3$]{%
    \includegraphics[scale=1.0]{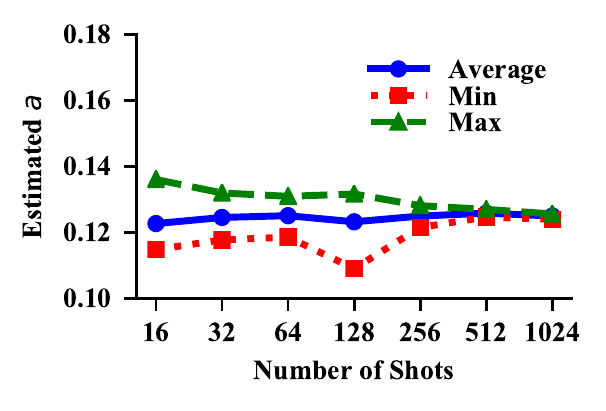}}
    \quad\quad\quad\quad\quad\quad\quad\quad
  \subfloat[$m=4$]{%
    \includegraphics[scale=1.0]{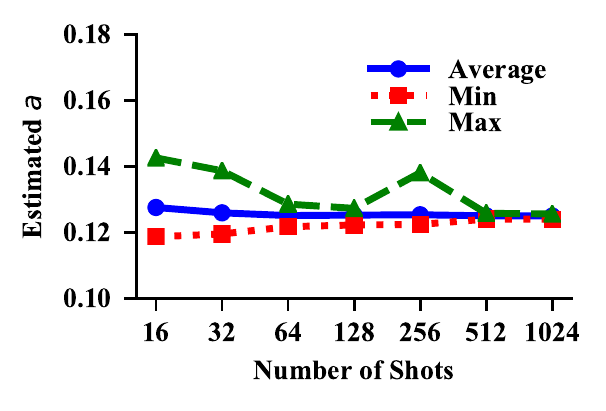}
  }\\
  \caption{Estimated $a$ by MLQAE on a $10$-qubit domain.}
  \label{fig:MLAE_a}
\end{figure*}

\begin{figure*}[t]
\centering
  \subfloat[$m=3$]{%
    \includegraphics[scale=1.0]{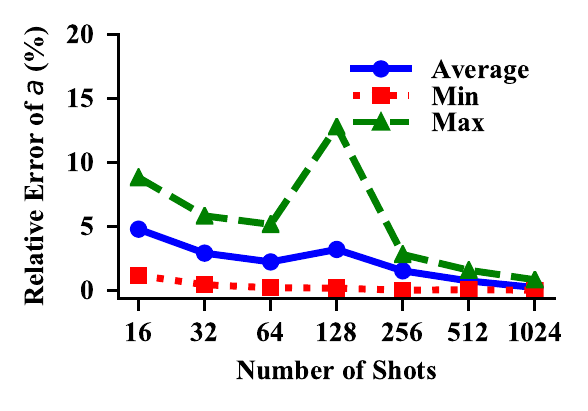}}
    \quad\quad\quad\quad\quad\quad\quad\quad
  \subfloat[$m=4$]{%
    \includegraphics[scale=1.0]{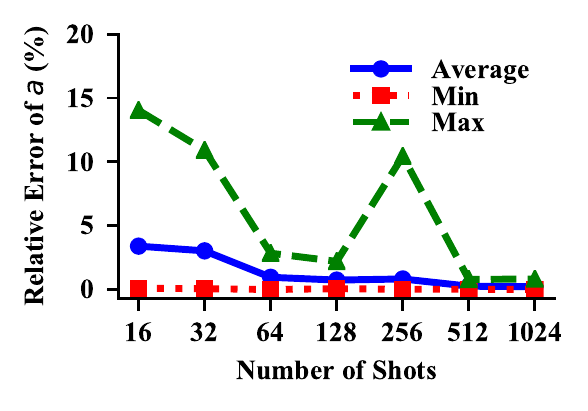}
  }\\
  \caption{The relative error of the estimated $a$ by MLQAE on a $10$-qubit domain.}\label{fig:MLAE_rel_a}
\end{figure*}

\begin{figure}[t]
\centering
    \includegraphics[scale=1.0]{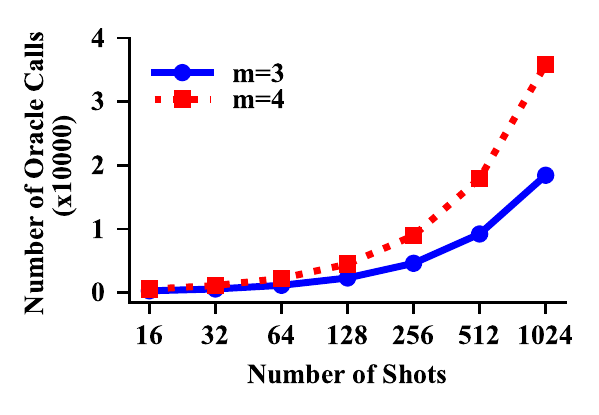}
  \caption{The number of oracle calls by MLQAE.}\label{fig:MLAE_oracle_num}
\end{figure}

\begin{figure*}[t]
\centering
  \subfloat[$m=3$]{%
    \includegraphics[scale=1.0]{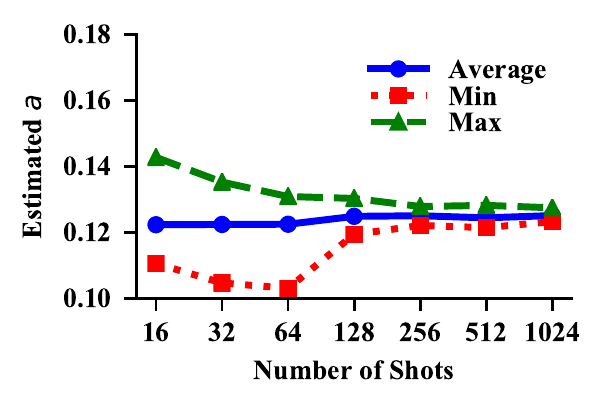}}
    \quad\quad\quad\quad\quad\quad\quad\quad
  \subfloat[$m=4$]{%
    \includegraphics[scale=1.0]{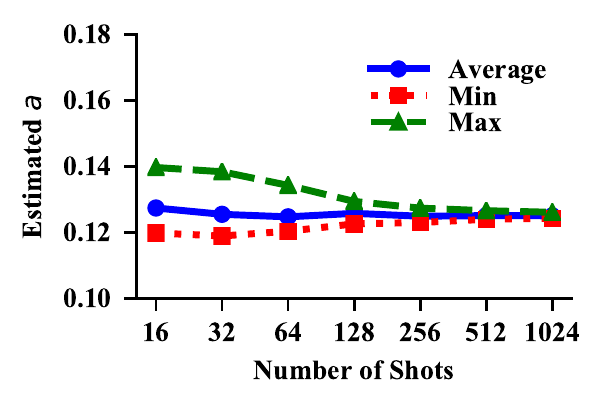}
  }\\
  \caption{Estimated $a$ by MLQAE on a $14$-qubit domain.}
  \label{fig:MLAE_a_14q}
\end{figure*}

\begin{figure*}[t]
\centering
  \subfloat[$m=3$]{%
    \includegraphics[scale=1.0]{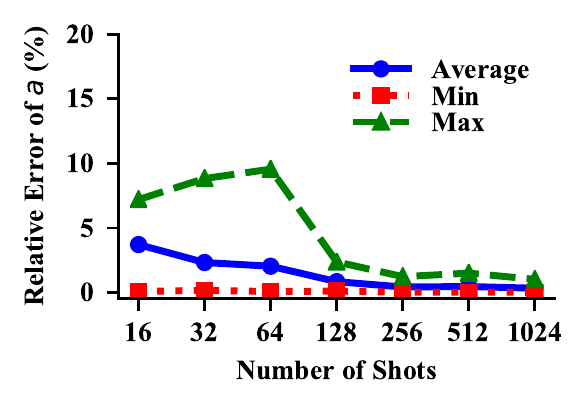}}
    \quad\quad\quad\quad\quad\quad\quad\quad
  \subfloat[$m=4$]{%
    \includegraphics[scale=1.0]{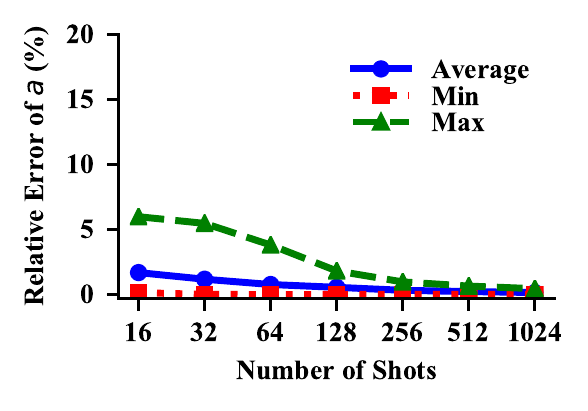}
  }\\
  \caption{The relative error of the estimated $a$ by MLQAE on a $14$-qubit domain.}
  \label{fig:MLAE_rel_a_14q}
\end{figure*}

\begin{figure*}[tb]
\centering
  \subfloat[Error bound $0.01$]{%
    \includegraphics[scale=1.0]{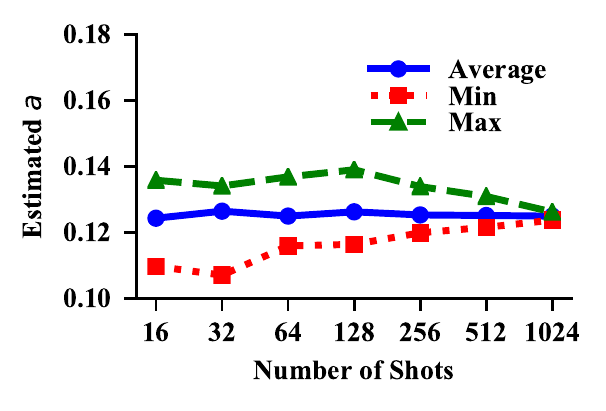}}
    \quad\quad\quad\quad\quad\quad\quad\quad
  \subfloat[Error bound $0.005$]{%
    \includegraphics[scale=1.0]{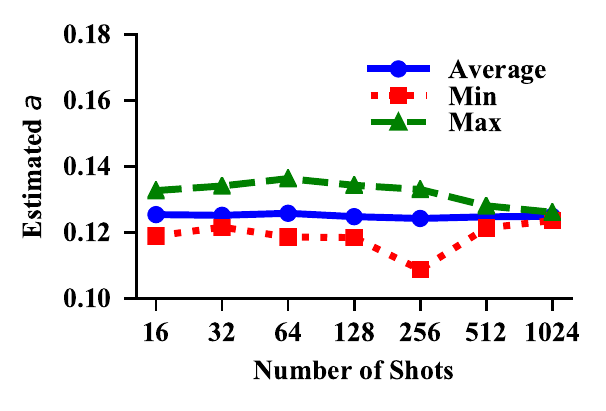}
  }\\
  \caption{Estimated $a$ by IQAE on a $10$-qubit domain.}
  \label{fig:IAE_a}
\end{figure*}

\begin{figure*}[tb]
\centering
  \subfloat[Error bound $0.01$]{%
    \includegraphics[scale=1.0]{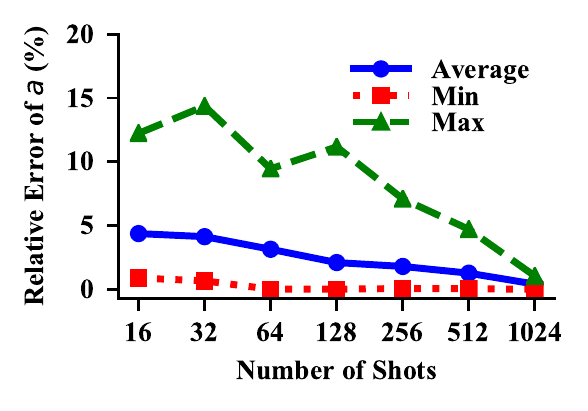}}
    \quad\quad\quad\quad\quad\quad\quad\quad
  \subfloat[Error bound $0.005$]{%
    \includegraphics[scale=1.0]{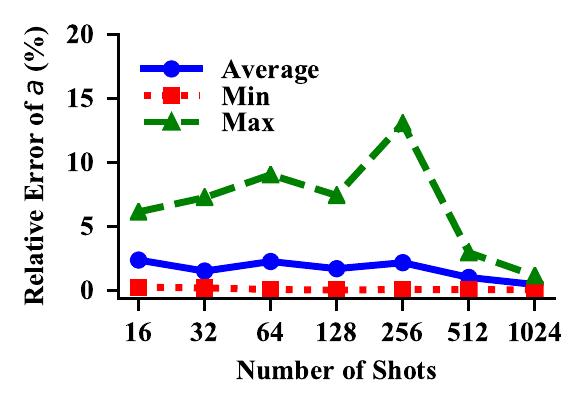}
  }\\
  \caption{The relative error of the estimated $a$ by IQAE on a $10$-qubit domain.}
  \label{fig:IAE_rel_a}
\end{figure*}

\begin{figure*}[tb]
\centering
  \subfloat[Error bound $0.01$]{%
    \includegraphics[scale=1.0]{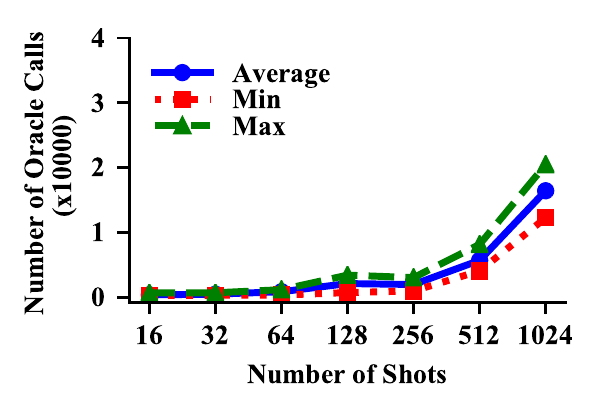}}
    \quad\quad\quad\quad\quad\quad\quad\quad
  \subfloat[Error bound $0.005$]{%
    \includegraphics[scale=1.0]{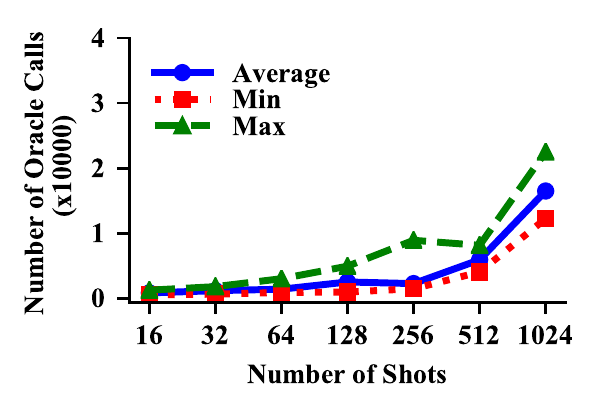}
  }\\
  \caption{The average of the number of oracle calls by IQAE on a $10$-qubit domain.}\label{fig:IAE_oracle_num_10q}
\end{figure*}

\begin{figure*}[tb]
\centering
  \subfloat[Error bound $0.01$]{%
    \includegraphics[scale=1.0]{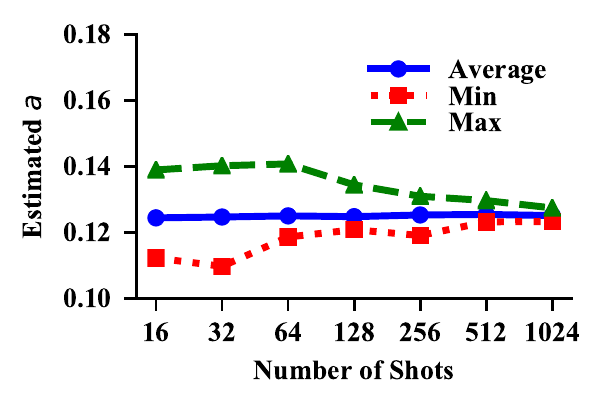}}
    \quad\quad\quad\quad\quad\quad\quad\quad
  \subfloat[Error bound $0.005$]{%
    \includegraphics[scale=1.0]{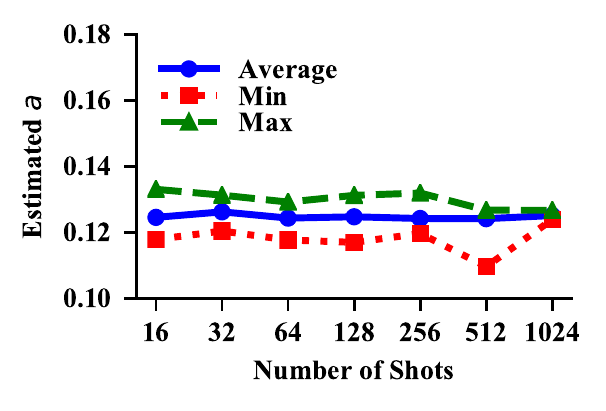}
  }\\
  \caption{Estimated $a$ by IQAE on a $14$-qubit domain.}\label{fig:IAE_a_14q}
\end{figure*}

\begin{figure*}[tb]
\centering
  \subfloat[Error bound $0.01$]{%
    \includegraphics[scale=1.0]{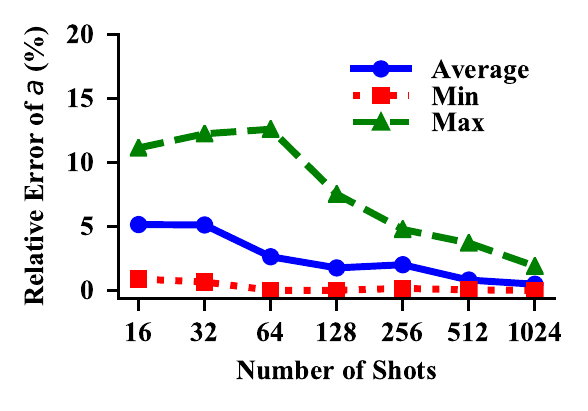}}
    \quad\quad\quad\quad\quad\quad\quad\quad
  \subfloat[Error bound $0.005$]{%
    \includegraphics[scale=1.0]{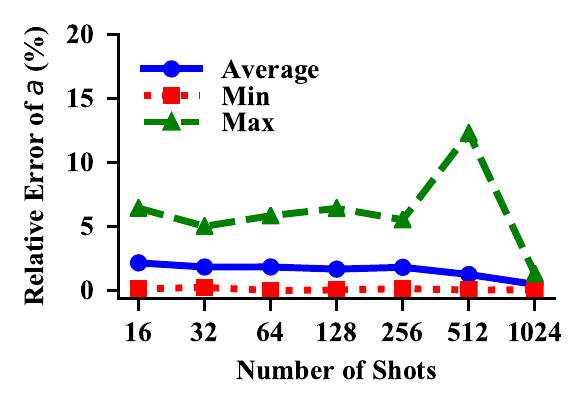}
  }\\
  \caption{The relative error of the estimated $a$ by IQAE on a $14$-qubit domain.}
  \label{fig:IAE_rel_a_14q}
\end{figure*}

\begin{figure*}[tb]
\centering
  \subfloat[Error bound $0.01$]{%
    \includegraphics[scale=1.0]{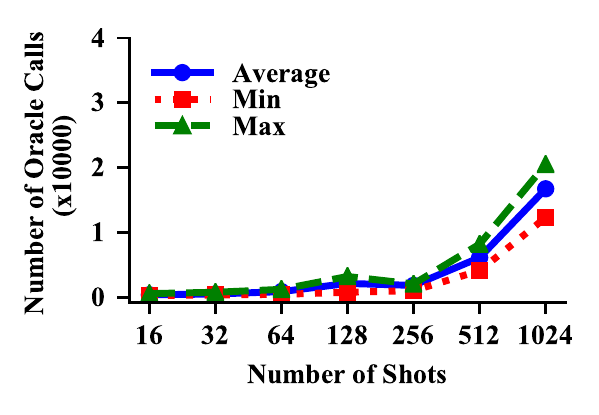}}
    \quad\quad\quad\quad\quad\quad\quad\quad
  \subfloat[Error bound $0.005$]{%
    \includegraphics[scale=1.0]{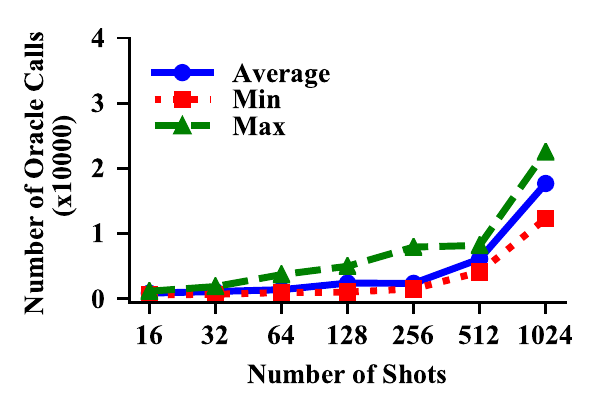}
  }\\
  \caption{The average of the number of oracle calls by IQAE on a 14-qubit domain.}
  \label{fig:IAE_oracle_num_14q}
\end{figure*}

In this section, we discuss the results obtained by running MLQAE and IQAE on IBMQ simulator.
MLQAE runs $m=3$ ($4$ circuits) and $m=4$ ($5$ circuits) cases.
IQAE runs with $0.01$ and $0.005$ set as error bound, $\epsilon$.
Different number of shots ($16, 32, \cdots, 1024$) in each case is tested for both MLQAE and IQAE.
In each case, the number of the oracle calls and estimated $\theta$ are recorded. From the estimated $\theta$, we can then calculate the estimated $a$.
Each case repeats $30$ times.
In all the simulations considered here, the true value of $a$ is $0.125~(=1/8)$.

In MLQAE, the number of the oracle calls is decided by $m$ and the number shots.
Since the operator $\textbf{Q}$ in Eq. (\ref{eq:qae_def_q}) includes $\mathcal{A}$ and $\mathcal{A}^{-1}$, the operator $\textbf{Q}$ has two oracle calls.
Therefore, when we have $\textbf{Q}^k \ket{\Psi}$ ($ = \textbf{Q}^k \mathcal{A} \ket{0}_{n+1}$), we have $2k + 1$ oracle calls.
For example, when we have $m=3$ case, the circuits are $\mathcal{A} \ket{0}_{n+1}$, $\textbf{Q} \mathcal{A} \ket{0}_{n+1}$, $\textbf{Q}^2 \mathcal{A} \ket{0}_{n+1}$, $\textbf{Q}^4 \mathcal{A} \ket{0}_{n+1}$, which yield $18$ as the cumulative number of oracle calls.
So, the total number of oracle calls in each of the simulations is $18$ times the number of shots.
When we have $m=4$ case, $\textbf{Q}^8 \mathcal{A} \ket{0}_{n+1}$ is added to the $m=3$ case, adding another $17$ oracle calls. 
Therefore, the total number oracle calls is $35$ times the number of shots.

In IQAE, the power $k$ of $\textbf{Q}$ is estimated for the circuit in each iteration.
Hence, there is no direct estimation of the oracle calls by the number of shots.
The number of oracle calls in each run is plotted for IQAE in the following subsection.
All data of the simulation results is listed in the Appendix.


\subsection{Maximum Likelihood Quantum Amplitude Estimation}



MLQAE runs with the two parameter values, namely with $m=3$ and $m=4$. These values are associated with the number of circuits (the likelihood functions Eq. (\ref{eq:lfunc})). The bigger $m$ is, the more accurate are the results. The estimated $a$ and its relative error by the maximum likelihood estimation for a $10$-qubit domain for $m=3$ and $m=4$ are presented in Figs.~\ref{fig:MLAE_a} and~\ref{fig:MLAE_rel_a}. 
As is the case with IQAE, for both the $10$-qubit and $14$-qubit domains in Figs.~\ref{fig:MLAE_a} and~\ref{fig:MLAE_a_14q}, the average values for estimated $a$ in MLQAE is stable in all range of shots.
The average relative error in Fig.~\ref{fig:MLAE_rel_a} is smaller for $m=4$ compared to $m=3$, almost half of the value of $m=3$ for smaller number of shots, and decreases at a faster rate than $m=3$ when the number of shots is increased. As expected, the maximum and the minimum values depart from the averages quite significantly in some cases. In Fig.~\ref{fig:MLAE_oracle_num}, there is a remarkable difference in the number of queries between $m=3$ and $m=4$, and it can be seen that the smaller relative error observed for $m=4$ comes at the cost of more oracle queries. Similar patterns for the average values of estimated $a$ and relative error of $a$ hold for the $14$-qubit data in Figs.~\ref{fig:MLAE_a_14q} and \ref{fig:MLAE_rel_a_14q}. Unlike IQAE, in MLQAE, the number of oracle queries do not change as the number of qubits in the domain is increased from $10$ to $14$ qubits. 

\subsection{Iterative Quantum Amplitude Estimation}

The estimated $a$ and its relative error by the iterative QAE (IQAE) for the $10$-qubit domain are shown in Figs.~\ref{fig:IAE_a} and~\ref{fig:IAE_rel_a}, respectively. In Fig.~\ref{fig:IAE_a}, the average, maximum and minimum values are plotted for the error bounds of $0.01$ and $0.005$. It can be observed that once the number of shots is large enough, the average value seems to converge for both of these cases, but there are some notable variations in the maximum and the minimum values. Additionally, the average values for $a$ for the two error tolerances considered here are comparable. However, the average of the relative error plotted in Fig.~\ref{fig:IAE_rel_a}(a) for $\epsilon=0.01$ is more sensitive to the number of shots, reducing dramatically as the number of shots is increased from $16$ to $1024$. In comparison, for $\epsilon=0.005$, the relative error is already quite small even for a smaller number of shots, and the overall reduction in error as the number of shots is increased is less pronounced in Fig.~\ref{fig:IAE_rel_a}(b). For the smaller error tolerance of $\epsilon=0.005$, the relative error for smaller number of shots, as low as $N=16$, is comparable to that of $N=128$ for $\epsilon=0.01$. 

One of the most crucial aspect of a quantum algorithm is its quantum complexity, the number of oracle queries required by it. For the $10$-qubit case, the average number of oracle queries in Fig.~\ref{fig:IAE_oracle_num_10q} for the two error tolerances increase with an increase in number of shots, but are close in magnitude. A similar theme ensues for the corresponding average values of $a$, relative error of $a$ and the number of oracle queries in the $14$-qubit domain shown in Figs.~\ref{fig:IAE_a_14q},~\ref{fig:IAE_rel_a_14q} and~\ref{fig:IAE_oracle_num_14q}.

\subsection{Comparison}
\label{sec:comparison}

\begin{figure}[t]
\centering
  \subfloat[1024 ($2^{10}$) samples]{%
    \includegraphics[scale=0.8]{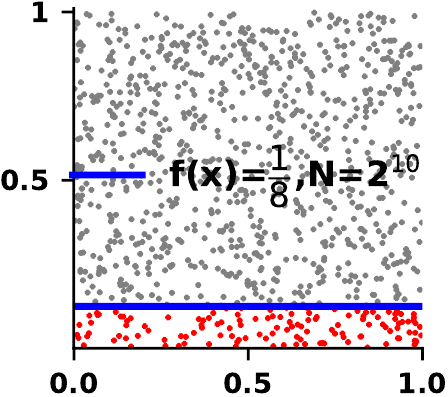}}
  \quad\quad
  \subfloat[16384 ($2^{14}$) samples]{%
    \includegraphics[scale=0.8]{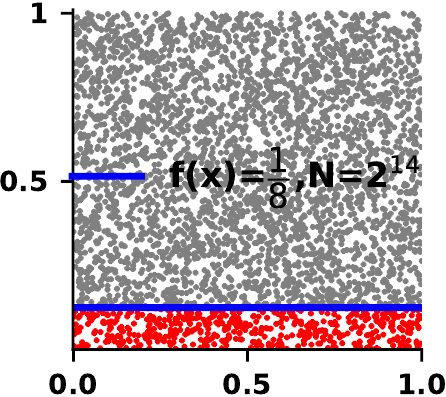}
  }\\
  \caption{The schematic diagram of Monte Carlo integration of function $f(x) = 1/8$.}\label{fig:MCI}
\end{figure}

\begin{figure*}[tb]
\centering
  \subfloat[$10$ qubits]{%
    \includegraphics[scale=1.0]{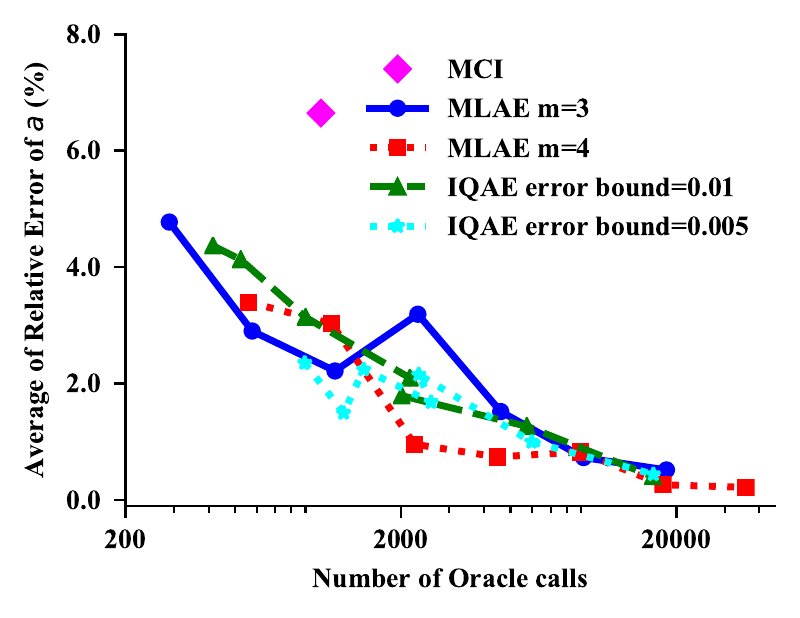}}
    \quad
  \subfloat[$14$ qubits]{%
    \includegraphics[scale=1.0]{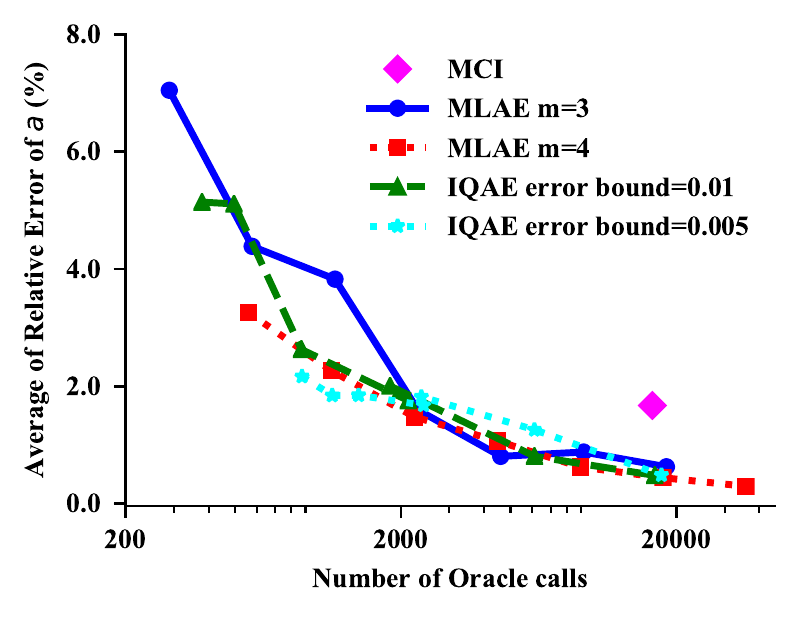}
  }\\
  \caption{The relative error of the estimated $a$ vs. the number of oracle calls.}\label{fig:ave_rel_vs_num_oracle}
\end{figure*}

We compare MLQAE and IQAE based on our simulation results.
These two quantum algorithms and implementations are also compared with the Monte Carlo integration (MCI).
Since the true value of $a$ is $1/8$ in the MLQAE and IQAE simulations, this is comparable to MCI of $f(x) = 1/8$ as discussed in~\cite{Yu2020Practical}.
The schematic diagram of MCI for $f(x) = 1/8$ is shown in Fig. \ref{fig:MCI}.
Since we test $10$-qubit and $14$-qubit domains in our QAE simulations, we use $2^{10}$ and $2^{14}$ samples in the MCI for facilitating direct comparison. 
Each case in the MCI repeats $10,000$ times.

One of the main points of comparison is the number of oracle calls.
In MCI, one oracle call computes the function value for each sample and decides whether the computed value is above (bad state) or below (good state) the threshold, as shown in Fig.~\ref{fig:MCI}. Thus, the number of oracle calls in MCI is equal to the number of samples.
Even though MLQAE and IQAE need additional post-processing by classical computing based on the results from quantum computing, we do not include the computing load from classical computing in our comparison because the load of the post-processing methods is constant with respect to the domain size (number of qubits in the domain).

Fig. \ref{fig:ave_rel_vs_num_oracle} shows averages of the relative errors with respect to the number of the oracle calls.
Neither MLQAE nor IQAE predominant over each other even though MLQAE with $m=4$ and IQAE with error bound $0.005$ are a little bit more accurate than MLQAE with $m=3$ and IQAE with error bound $0.005$, respectively.
In 10-qubit domain, the quantum methods are 2 and more times accurate than the classical MCI.
In 14-qubit domain, the quantum methods are 2.5 and more times accurate than the classical MCI.

%% file: conclusion.tex
\section{Conclusion}
\label{sec:conclusion}


In this paper, we have implemented and discussed recently developed two quantum amplitude estimation algorithms on IBM quantum simulator, in the broader context of Monte Carlo integration.
While MLQAE and IQAE achieve comparable accuracy in similar number of oracle calls, they both predominate the classical MCI in accuracy with respect to the number of oracle calls.

Even though MLQAE and IQAE are comparable in accuracy and the number of oracle calls, they have contrasting properties.
IQAE is controlled by specifying an error bound while MLQAE cannot control the error bound directly and in fact its upper error bound seems unclear.
Being able to control the error offers IQAE a very significant advantage over MLQAE.
On the other hand, the circuit depth of IQAE is not controllable because it sometimes has very large powers of $\textbf{Q}$ (equivalently many oracle calls) as shown in Figs.~\ref{fig:IAE_oracle_num_10q} and~\ref{fig:IAE_oracle_num_14q} while MLQAE can control its circuit depth by choosing the number of shots and the value of $m$.
This is a critical disadvantage of IQAE on NISQ devices since circuit depth is strictly limited due to gate errors and decoherence on these devices.

%% file: appendix.tex


$\newline$

\appendix


\begin{table}[h]
\caption{Monte Carlo integration}
\label{table:MCI_1024}
\begin{center}
\begin{tabular}{l | c  c }
    \hline \hline
    Sample Size & 1024 & 16384  \\ \hline
    Max of Integral & 0.169 & 0.135	\\
    Avg of Integral & 0.125	& 0.125 \\
    Min of Integral & 0.084	& 0.114 \\
    Std Dev of Integral & 0.010 & 0.003\\
    
    Max Rel Err (\%) & 35.157 & 8.594 \\
    Avg Rel Err (\%) & 6.645 & 1.673	 \\
    Min Rel Err (\%) & 0.0 	& 0.0	\\
    Std Dev Rel Err & 5.002 & 1.257\\
    \hline \hline
\end{tabular}
\end{center}
\end{table}

\begin{table*}[ht]
\caption{MLQAE data for a $10$-qubit domain for $m=3$.}
\label{table:mlae_10q_3m}
\begin{center}
\begin{tabular}{l | c  c  c c c c c}
    \hline \hline
    Shots & 16 & 32 & 64 & 128 & 256 & 512 & 1024 \\ \hline
    Max $a$ & 0.136 & 0.132 & 0.131 & 0.132 & 0.128 & 0.127 & 0.126\\
    Avg $a$ & 0.123 & 0.125 & 0.125 & 0.123 & 0.125 & 0.126 & 1.125  \\
    Min $a$ & 0.115 & 0.118 & 0.119 & 0.119 & 0.122 & 0.125 & 0.124  \\
    Std Dev $a$ &0.007 & 0.004 & 0.004 & 0.005 & 0.002 & 0.001 & 0.001 \\
    
    Max Err & 8.801 & 5.812 & 5.146 & 12.775 & 2.794 & 1.556 & 0.872  \\
    Avg Err & 2.507 & 1.520 & 1.159 & 1.685 & 0.794 & 0.380 & 0.269 \\
    Min Err & 0.609 & 0.227 & 0.103 & 0.088 & 0.001 & 0.035 & 0.001 \\
    Std Dev Err & 2.817	& 1.648 & 1.672	& 3.188	& 0.939	& 0.498	&0.268 \\

    \hline \hline
\end{tabular}
\end{center}
\end{table*}

\begin{table*}[ht]
\caption{MLQAE data for a $10$-qubit domain for $m=4$.}
\label{table:mlae_10q_4m} 
\begin{center}
\begin{tabular}{l | c  c  c c c c c}
    \hline \hline
    Shots & 16 & 32 & 64 & 128 & 256 & 512 & 1024 \\ \hline
    Max $a$ & 0.143	& 0.139	& 0.129	& 0.127	& 0.138	& 0.126 & 0.126 \\
    Avg $a$ & 0.128 & 0.126	& 0.125	& 0.125 & 0.125 & 0.125 & 0.125 \\
    Min $a$ & 0.119	& 0.120	& 0.122	& 0.122	& 0.122	& 0.124 & 0.124 \\
    Std Dev $a$ & 0.005	& 0.006	& 0.002	& 0.001	& 0.003	& 0.000	& 0.000 \\
    
    Max Err& 14.039 & 10.912 & 2.841 & 2.202 & 10.443 & 0.789 & 0.822 \\
    Avg Err& 3.397 & 3.027 & 0.956 & 0.738 & 0.826 & 0.260 & 0.216 \\
    Min Err& 0.098 & 0.065 & 0.002 & 0.065 & 0.018 & 0.018 & 0.006 \\
    Std Dev Err & 3.240	& 3.249	& 0.788	& 0.554	& 1.877	& 0.220	& 0.190 \\
    \hline \hline
\end{tabular}
\end{center}
\end{table*}

\begin{table*}[ht]
\caption{MLQAE data for a $14$-qubit domain for $m=3$.}
\label{table:mlae_14q_3m}
\begin{center}
\begin{tabular}{l | c  c  c c c c c}
    \hline \hline
    Shots & 16 & 32 & 64 & 128 & 256 & 512 & 1024 \\ \hline
    Max $a$ & 0.143 & 0.135 & 0.131 & 0.130 & 0.128 & 0.128 & 0.127	\\
    Avg $a$ & 0.122	& 0.122	& 0.122	& 0.125	& 0.125	& 0.124	& 0.125 \\
    Min $a$ & 0.111	& 0.105	& 0.103	& 0.119	& 0.122	& 0.121	& 0.123 \\
    Std Dev $a$ & 0.010	& 0.007	& 0.007	& 0.003	& 0.001	& 0.001	& 0.001  \\
    
    Max Err& 14.162	& 16.253 & 17.526 & 4.479 & 2.334 & 2.843 & 1.951 \\
    Avg Err& 7.044	& 4.384	& 3.825	& 1.598	& 0.805	& 0.879	& 0.628 \\
    Min Err& 0.148	& 0.317	& 0.131	& 0.201	& 0.035	& 0.035	&0.026 \\
    Std Dev Err & 3.602	& 3.537	& 4.107	& 1.237	& 0.653	& 0.770	& 0.443  \\
    \hline \hline
\end{tabular}
\end{center}
\end{table*}

\begin{table*}[ht]
\caption{MLQAE data for a $14$-qubit domain for $m=4$.}
\label{table:mlae_14q_4m}
\begin{center}
\begin{tabular}{l | c  c  c c c c c}
    \hline \hline
    Shots & 16 & 32 & 64 & 128 & 256 & 512 & 1024 \\ \hline
    Max $a$ & 0.140 & 0.138 & 0.134 & 0.129 & 0.127 & 0.127 & 0.126\\
    Avg $a$ & 0.127 & 0.126 & 0.125 & 0.126 & 0.125 & 0.125 & 0.125\\
    Min $a$ & 0.120 & 0.119 & 0.120 & 0.123 & 0.123 & 0.124 & 0.124 \\
    Std Dev $a$ & 0.005	& 0.004	& 0.003	& 0.002	& 0.001	& 0.001	& 0.000  \\
    
    Max Err& 11.712 & 10.686 & 7.392 & 3.498 & 1.868 & 1.248 & 0.864\\
    Avg Err& 3.256 & 2.263 & 1.477 & 1.066 & 0.625 & 0.440 & 0.295 \\
    Min Err& 0.281	& 0.032 & 0.002 & 0.018 & 0.015	& 0.044 & 0.006 \\
    Std Dev Err & 2.701	& 2.624	& 1.510	& 1.024	& 0.526	& 0.283	& 0.215  \\
    \hline \hline
\end{tabular}
\end{center}
\end{table*}

\begin{table*}[ht]
\caption{IQAE data for a $10$-qubit domain for $\epsilon=0.01$.}
\label{table:iae_10q_0.01}
\begin{center}
\begin{tabular}{l | c  c  c c c c c}
    \hline \hline
    Shots & 16 & 32 & 64 & 128 & 256 & 512 & 1024 \\ \hline
    Max $a$ & 0.136 & 0.134 & 0.137 & 0.139 & 0.134 & 0.131 & 0.126	\\
    Avg $a$ & 0.124 & 0.126 & 0.125 & 0.126 & 0.125 & 0.125 & 0.125 \\
    Min $a$ & 0.110 & 0.107 & 0.116 & 0.116 & 0.120 & 0.122 & 0.124\\
    Std Dev $a$ & 0.007	& 0.006	& 0.005	& 0.004	& 0.003	& 0.002	& 0.001  \\
    
    Max Err& 12.224	& 14.362 & 9.446 & 11.165 & 7.090 & 4.701 & 1.036 \\
    Avg Err& 4.367 & 4.127 & 3.137 & 2.095 & 1.790 & 1.264 & 0.403  \\
    Min Err& 0.883	& 0.658	& 0.005	& 0.008	& 0.060	& 0.046	& 0.000 \\
    Std Dev Err & 3.188	& 2.711	& 2.146	& 2.892	& 1.541	& 1.041	& 0.292  \\
    
    Max Oracle Call & 720 & 736	& 1216 & 3456 & 3072 & 8192 & 20480 \\
    Avg Oracle Call & 414.933 & 523.733	& 900.267 & 2154.267 & 2013.867 & 5734.4 & 16452.267 \\
    Min Oracle Call & 304 & 352	& 384 & 768	& 1024 & 4096 & 12288 \\
    Std Dev Ora Call & 99.781 & 108.430	& 177.818 & 916.931	& 501.896 & 954.556	& 2434.144 \\
    \hline \hline
\end{tabular}
\end{center}
\end{table*}

\begin{table*}[ht]
\caption{IQAE data for a $10$-qubit domain for $\epsilon=0.005$.}
\label{table:iae_10q_0.005}
\begin{center}
\begin{tabular}{l | c  c  c c c c c}
    \hline \hline
    Shots & 16 & 32 & 64 & 128 & 256 & 512 & 1024 \\ \hline
    Max $a$ & 0.133 & 0.134 & 0.136 & 0.134 & 0.133 & 0.128 & 0.126	\\
    Avg $a$ & 0.125 & 0.125 & 0.126 & 0.125 & 0.124 & 0.125 & 0.125	\\
    Min $a$ & 0.119 & 0.122 & 0.119 & 0.118 & 0.109 & 0.121 & 0.124 \\
    Std Dev $a$ & 0.004	& 0.003	& 0.004	& 0.003	& 0.004	& 0.002	& 0.001  \\
    
    Max Err& 6.119 & 7.238 & 9.021 & 7.401 & 13.027	& 2.937	& 1.129 \\
    Avg Err& 2.355 & 1.494 & 2.242 & 1.680 & 2.152 & 0.997 & 0.441 \\
    Min Err& 0.233 & 0.181 & 0.058 & 0.008 & 0.063 & 0.041 & 0.035\\
    Std Dev Err & 1.765	& 1.324	& 2.246	& 1.753	& 2.644	& 0.787	& 0.296  \\
    
    Max Oracle Call & 1312 & 1856 & 3072 & 4992	& 8960 & 8192 & 22528 \\
    Avg Oracle Call & 894.4 & 1238.4 & 1463.467	& 2572.8 & 2321.067 & 5973.333 & 16521.533 \\
    Min Oracle Call & 576 & 704 & 960 & 1024 & 1536	& 4096 & 12288 \\
    Std Dev Ora Call & 221.597 & 370.250 & 627.685 & 1112.506 & 1587.988 & 1263.712	& 3178.820  \\
    \hline \hline
\end{tabular}
\end{center}
\end{table*}

\begin{table*}[ht]
\caption{IQAE data for a $14$-qubit domain for $\epsilon=0.01$.}
\label{table:iae_14q_0.01}
\begin{center}
\begin{tabular}{l | c  c  c c c c c}
    \hline \hline
    Shots & 16 & 32 & 64 & 128 & 256 & 512 & 1024 \\ \hline
    Max $a$ & 0.139 & 0.140 & 0.141 & 0.134 & 0.1317 & 0.130 & 0.127\\
    Avg $a$ & 0.124 & 0.125 & 0.125 & 0.125 & 0.125 & 0.125 & 0.125 \\
    Min $a$ & 0.112 & 0.110 & 0.119 & 0.121 & 0.119 & 0.123 & 0.123\\
    Std Dev $a$ & 0.008	& 0.008	& 0.004	& 0.003	& 0.003	& 0.001	& 0.001 \\
    
    Max Err& 11.123 & 12.224 & 12.580 & 7.513 & 4.759 & 3.699 & 1.893 \\
    Avg Err& 5.138 & 5.108 & 2.626 & 1.754 & 2.004 & 0.807 & 0.480 \\
    Min Err& 0.883 & 0.658 & 0.005 & 0.008 & 0.156 & 0.041 & 0.000 \\
    Std Dev Err & 3.453	& 3.650	& 2.369	& 1.604	& 1.312	& 0.904	& 0.434 \\
    
    Max Oracle Call & 560 & 768	& 1216 & 3200 & 2048 & 8192	& 20480 \\
    Avg Oracle Call & 378.667 & 494.933	& 870.400 & 2129.067 & 1826.133 & 6109.867 & 16725.333 \\
    Min Oracle Call & 256 & 352	& 448 & 768	& 1024 & 4096 & 12288 \\
    Std Dev Ora Call & 80.751 & 129.844	& 207.622 & 887.118	& 347.602 & 1303.154 & 3093.503  \\
    \hline \hline
\end{tabular}
\end{center}
\end{table*}

\begin{table*}[ht]
\caption{IQAE data for a $14$-qubit domain for $\epsilon=0.005$.}
\label{table:iae_14q_0.005}
\begin{center}
\begin{tabular}{l | c  c  c c c c c}
    \hline \hline
    Shots & 16 & 32 & 64 & 128 & 256 & 512 & 1024 \\ \hline
    Max $a$ & 0.133 & 0.131 & 0.129 & 0.131 & 0.132 & 0.127 & 0.127	\\
    Avg $a$ & 0.125 & 0.126 & 0.124 & 0.125 & 0.124 & 0.124 & 0.125 \\
    Min $a$ & 0.118 & 0.120 & 0.118 & 0.117 & 0.120 & 0.110 & 0.124 \\
    Std Dev $a$ & 0.004	& 0.003	& 0.003	& 0.003	& 0.003	& 0.003	& 0.001  \\
    
    Max Err (\%)& 6.430 & 5.017 & 5.844 & 6.419 & 5.528 & 12.275 & 1.320 \\
    Avg Err (\%)& 2.167 & 1.844 & 1.847 & 1.677 & 1.819 & 1.259 & 0.486 \\
    Min Err (\%)& 0.137 & 0.251 & 0.005 & 0.056 & 0.157 & 0.046 & 0.063 \\
    Std Dev Err & 1.769	& 1.460	& 1.503	& 1.723	& 1.327	& 2.157	& 0.290 \\
    
    Max Oracle Call& 1168 & 1888 & 3712 & 4992	& 7936 & 8192 & 22528 \\
    Avg Oracle Call& 874.133 & 1125.333 & 1403.733 & 2414.933 & 2372.267 & 6109.867 & 17681.067 \\
    Min Oracle Call& 592 & 704	& 960 & 1024 & 1536 & 4096 & 12288 \\
    Std Dev Ora Call & 180.310 & 311.397 & 698.447 & 1134.822 & 1430.781 & 1435.196	& 3292.798  \\
    \hline \hline
\end{tabular}
\end{center}
\end{table*}